\documentclass[]{spie}  %>>> use for US letter paper
\usepackage[latin1]{inputenc}
\usepackage[british]{babel}
\usepackage[]{graphicx}
\usepackage{amsmath,amssymb}
\usepackage{textcmds}
\usepackage{astro_bib_macro} %raccourcis noms de journaux d?finis dans ce fichier
\usepackage{multirow}
\usepackage{color}

\title{High-contrast imager for Complex Aperture Telescopes\\ (HiCAT): 2. design overview and first light results} 

\author{
Mamadou N'Diaye\supit{a}*, 
Elodie Choquet\supit{a},
Sylvain Egron\supit{b,a},
Laurent Pueyo\supit{a,c},\\
Lucie Leboulleux\supit{b,a},
Olivier Levecq\supit{b,a},
Marshall D. Perrin\supit{a},
Erin Elliot\supit{a},
J. Kent Wallace\supit{d},\\
Emmanuel Hugot\supit{e},
Michel Marcos\supit{e},
Marc Ferrari\supit{e}, 
Chris A. Long\supit{a},
Rachel Anderson\supit{a},\\
Audrey DiFelice\supit{a},
and R\'emi Soummer\supit{a\dag}
\skiplinehalf
\supit{a} Space Telescope Science Institute, 3700 San Martin Drive, Baltimore, MD 21218, USA\\
\supit{b} Institut d'Optique Graduate School (Palaiseau, Saint-Etienne, Bordeaux), France\\
\supit{c} Dept. of Physics and Astronomy, Johns Hopkins University, Baltimore, MD 21218, USA\\
\supit{d} Jet Propulsion Laboratory, California Institute of Technology, Pasadena, CA 91109, USA\\%4800 Oak Grove Drive, 
\supit{e} Aix Marseille Universit\'e, CNRS, LAM UMR 7326, 13388 Marseille, France\\
}

\authorinfo{*E-mail: mamadou@stsci.edu, \dag: Principal Investigator}
\begin{document} 
  \maketitle 

%%%%%%%%%%%%%%%%%%%%%%%%%%%%%%%%%%%%%%%%%%%%%%%%%%%%%%%%%%%%% 
\begin{abstract}
We present a new high-contrast imaging testbed designed to provide complete solutions in wavefront sensing, control and starlight suppression with complex aperture telescopes. The testbed was designed to enable a wide range of studies of the effects of such telescope geometries, with primary mirror segmentation, central obstruction, and spiders. The associated diffraction features in the point spread function make high-contrast imaging more challenging. In particular the testbed will be compatible with both AFTA-like and ATLAST-like aperture shapes, respectively on-axis monolithic, and on-axis segmented telescopes. The testbed optical design was developed using a novel approach to define the layout and surface error requirements to minimize amplitude-induced errors at the target contrast level performance. In this communication we compare the as-built surface errors for each optic to their specifications based on end-to-end Fresnel modelling of the testbed. We also report on the testbed optical and optomechanical alignment performance, coronagraph design and manufacturing, and preliminary first light results. 
\end{abstract}

%>>>> Include a list of keywords after the abstract 


%%%%%%%%%%%%%%%%%%%%%%%%%%%%%%%%%%%%%%%%%%%%%%%%%%%%%%%%%%%%%
\section{INTRODUCTION}\label{sec:intro}
The high-contrast imaging instruments Palomar/P1640\cite{2011PASP..123...74H}, Gemini/GPI\cite{2014arXiv1403.7520M} VLT/SPHERE\cite{2008SPIE.7014E..41B} have recently been deployed on ground-based telescopes to directly image and characterize young or massive hot, gaseous planets around nearby stars. These planetary companions will present contrast ratios up to $10^{7}$ in the near infrared. 
By 2018, the 6.5m James Webb Space Telescope \cite{2010ASPC..430..167C}, will be launched with instruments offering high-contrast imaging capabilities on both MIRI and NIRCam instruments and will enable observations of these types of objects at longer wavelengths. 

The search for fainter companions such as habitable terrestrial worlds by direct imaging (up to three orders of magnitude fainter) will require a telescope large enough to provide angular resolution and sensitivity to planets around a significant number of nearby stars. Segmented telescopes constitute a compelling option to address these questions. However, their aperture geometry (segmentation, central obstruction and spider vanes) makes high-contrast imaging more challenging. We are developing a testbed at the Space Telescope Science Institute (STScI) to provide an integrated solution for wavefront sensing and control and starlight suppression strategies on such complex aperture geometries. 

This communication reports on the development and current status of the High-contrast Imager for Complex Aperture Telescopes (HiCAT). The testbed was designed to minimize the impact of its optical surfaces on its contrast performance, by setting a contrast floor from the optical design and surface quality of the optics to below $10^{-8}$ assuming a single deformable mirror (DM), and even lower with two DMs as equiped on HiCAT. A theoretical contrast floor at this level is sufficiently below our contrast expectations for the success of HiCAT in the presence of complex aperture features (central obstruction, spiders and segments), and considering broadband operations in air. 

In Section \ref{sec:optimization}, we review our methodology based on an hybrid approach that combines optical ray tracing and end-to-end simulations. This approach allowed us to derive the final surface error specifications used for optics manufacturing. The as-built results are presented in Section \ref{sec:optics_design} and compared with the original surface error specifications. We give an overview of the testbed opto-mechanical design to underline its main features in Section \ref{sec:optomech}. Testbed alignment has been performed over the first semester 2014, resulting in the first light results that are exposed in Section \ref{sec:first_light}. As a conclusion, we discuss the next steps of the project.

%%%%%%%%%%%%%%%%%%%%%%%%%%%%%%%%%%%%%%%%%%%%%%%%%%%%%%%%%%%%%
\section{APPROACH FOR DESIGN DEVELOPMENT}\label{sec:optimization}
This section presents our methodology for the development of the testbed optical design. We briefly list the different constraints and describe our iterative approach. More details can be found in paper I\cite{2013SPIE.8864E..1KN}. 

\subsection{Optical design description}

\begin{figure}[!ht]
\centering
\includegraphics[height=5cm]{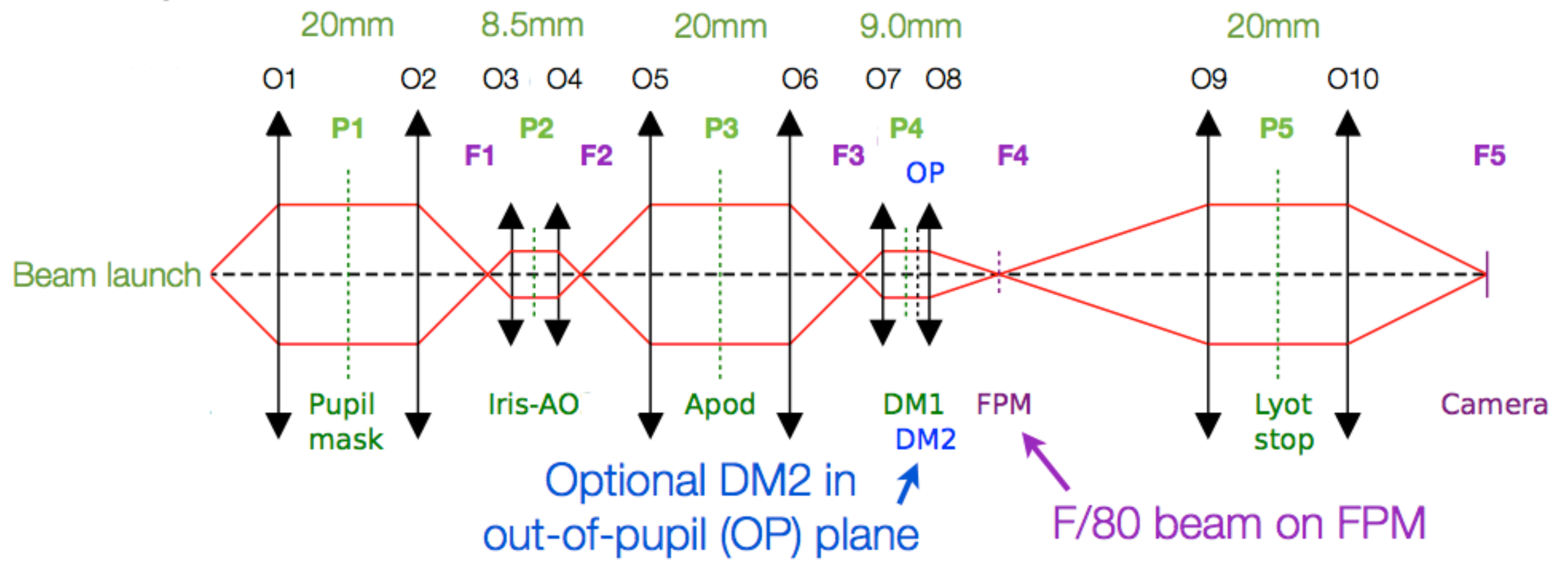}
\caption{Schematic linear representation of the high-contrast imaging testbed (light propagation goes from the left to the right). This layout involves 5 pupil planes (P), 5 focal planes (F) and 1 out-of-pupil (OP) plane for a second deformable mirror, leading to ten optics (O) for a design with collimated beams. The beam size in each pupil plane is indicated on top. The apodizer (Apod), focal plane mask (FPM), and Lyot stop constitutes the components of the APLC architecture-based coronagraphic system. The terms Iris-AO and DM refer to the segmented and the face-sheet deformable mirrors. The actual testbed implementation is fully reflective with mirrors to ease broadband operations, except O10 for which a refractive lens is used to image the field on the focal plane camera. The choice for the latter optic relaxes the length of the footprint of the design and leaves room for the pupil imager.}
\label{fig:testbed_conceptual_design}
\end{figure}

A conceptual optical layout of HiCAT is given in Figure \ref{fig:testbed_conceptual_design}. This layout was established to achieve HiCAT's goals: it includes 5 pupil planes (pupil mask to define central obstruction and support structures, Iris AO DM, Boston Micromachines DM, Apodizer and Lyot stop), and 10 powered optics. Because of our broadband goals the testbed was designed as purely reflective (including the apodizer and focal plane mask) up to the final optic O10 (located after the coronagraph), which is transmissive. 

The telescope aperture is defined by a 20\,mm entrance pupil mask with central obstruction and spiders and a 37 hexagonal segment Iris AO\footnote{http://www.irisao.com} MEMs deformable mirror that can be used in a conjugated plane to provide a segmented pupil similar to ATLAST. The DM can be interchanged with a high-quality flat mirror to enable monolithic telescope geometries such as AFTA. 

The diffraction suppression system is based on the architecture of the Apodized Pupil Lyot Coronagraph (APLC)\cite{2002A&A...389..334A,2005ApJ...618L.161S,2009ApJ...695..695S,2011ApJ...729..144S} for its performance and flexibility of design and operations (e.g. compatible with low order wavefront sensors\cite{2011SPIE.8126E..11W,2012SPIE.8450E..0NN,2013A&A...555A..94N}. Our design takes advantage of an existing focal plane mask (FPM) formerly used on the Lyot Project Coronagraph \cite{2004SPIE.5490..433O}. We are currently investigating novel APLC types of solutions, which can deliver $10^{-8}$ contrast in 20\% broadband, with an IWA of $\sim4\lambda/D$ and in the presence of $20\%$ central obstruction. These new types of solutions are presented in a companion paper by N'Diaye et al. in these proceedings\cite{2014ndiaye..aplc}.

Two Boston Micromachines\footnote{http://www.bostonmicromachines.com} deformable mirrors (kilo-DM) are included for phase and amplitude control applied to both the correction of the residual aberrations and to wavefront shaping (generation of a dark hole inside the point spread function [PSF]). These DMs will also be used to implement the novel technique Active Correction of Aperture Discontinuities (ACAD)\cite{2013ApJ...769..102P}. 

Two cameras (CamF and CamP) respectively for focal and pupil plane imaging are part of the back end. CamF is positioned on a translation stage to enable direct and coronagraphic phase diversity measurements\cite{2003JOSAA..20.1490D,2012OptL...37.4808S}.

Additional features have been identified outside this main optical path, including a 4D AccuFiz interferometer for alignment and direct wavefront sensing measurements, additional room for future implementation of low- and high-order wavefront sensing concepts, such as Zernike wavefront sensors \cite{2011SPIE.8126E..11W,2012SPIE.8450E..0NN,2013A&A...555A..94N}.

%_____________________________________________________________
%_____________________________________________________________

\subsection{Design methodology} 

Our design was established to set a contrast floor due to the surface quality of bench optics and propagation effects better than $10^{-8}$ in $2\%$ broadband using a single continuous deformable mirror (much lower with two DMs). This contrast floor is significantly below the contrast expectations for HiCAT,  and therefore guarantees that the testbed's performance will not be limited by the mirrors surface quality and associated propagation effects, but indeed by the effect we are interested in studying (broadband high-contrast for on-axis / segmented geometries). One important motivation for this approach is to ensure sufficient stroke availability to correct the aperture geometric features in large spectral broadband with two DMs. 

More specifically for the design, contrast must not be limited by amplitude-induced errors from the propagation of out-of-pupil surfaces (aka Talbot effects). To achieve this goal, we developed a new approach based on a minimization of the amplitude induced-errors from these Talbot effects. 
Indeed we placed a requirement on the contrast contribution from Talbot-type amplitude errors to define a comfortable contrast floor of $10^{-8}$ inside a half-field dark zone produce by a single DM from 3 to 10\,$\lambda/D$ with a 2\% bandpass. We adopted an hybrid approach combining analytical model, optics ray tracing and end-to-end Fresnel simulations to define the testbed design and determine the mirror specifications. 

Our first step included an optimization of the geometric optics design, with optical ray tracing in Zemax, opto-mechanical considerations and practical constraints to fit the testbed design on the table with sufficient clearance for each component, and an analytical optimization of the Talbot distances to minimize the Talbot effects. As a second step, we proceed to end-to-end testbed simulations with Fresnel propagation and starlight suppression algorithm to minimize amplitude-induced errors, building on previous analysis\cite{2006ApOpt..45.5143S,2007ApJ...666..609P,2010SPIE.7736E.109A}.  

We performed several iterations for both the geometric/Talbot distances and end-to-end optimizations to obtain an optical design with optics specifications that mitigate the Talbot effects under the opto-mechanical requirements. The details of this work and results were detailed in Paper I\cite{2013SPIE.8864E..1KN}.

%%%%%%%%%%%%%%%%%%%%%%%%%%%%%%%%%%%%%%%%%%%%%%%%%%%%%%%%%%%%%
\section{Optics design: specifications and as-built results}\label{sec:optics_design}

Using the end-to-end Fresnel simulation results of the geometrical optical layout, we established final requirements for each optics both in terms of surface quality and alignment tolerances. 

As a starting point for the design and its associated wavefront error (WFE) budget, we assumed that the contribution from each optic would be allocated evenly between the wavefront errors $\sigma_{WFE}$ from surface quality (mainly set by low-order WFE), and alignment contribution $\sigma_{Align}$. We assumed $\lambda/20$ rms each at He-Ne (i.e. $\lambda/40$ rms surface error for each mirror), corresponding to a total of 46 nm rms per optic (including both surface and alignment budget). 
This alignment error budget translates into comfortable mechanical tolerances (typically in the range 100 to 500\,$\mu$m alignment tolerances depending on the optic and based on our more detailed alignment tolerancing study with Zemax). 

We further allocated the WFE essentially to the low-order specification with additional very small mid- and high-spatial frequencies. The acceptable level for mid- and high-spatial frequency was set by the combination of Fresnel simulations and stroke minimization algorithm\cite{Pueyo:09} to deliver a contrast floor in $2\%$ broadband $\sim 10^{-8}$. The Fresnel simulation was designed to simulate the effect of a pre-compensation of the phase using the first DM to reduce the phase error into the linear regime for the stroke minimization algorithm\cite{Pueyo:09}. This is why we can tolerate an apparently-high total WFE (a direct RSS adds up to 130\,nm rms prior to correction by the DM, which would not be in the linear regime). This result underlines the importance of this simulation approach to relax the specifications to minimize cost as much as possible since pure phase can be corrected with the DM, while making sure that the experiment is never limited by propagation-induced amplitude errors. 
Table \ref{table:optics_specs_for_manufacturers} details all these specifications for each optic (surface error for low/mid/high spatial frequencies) and alignment budget. 

These simulations account for the fact that the optics are significantly oversized (a Fresnel propagation requirement) so that the actual WFE inside the pupil aperture is significantly smaller than for the full optic. Moreover we verified that these specifications on the optics included a comfortable margin (here about a factor 2 on the WFE) to guarantee convergence of the stroke minimization algorithm to our target contrast floor in $2\%$ broadband, with a single DM phase pre-compensation. 

In order to obtain a more realistic error budget, we assumed that most of the low-order aberrations (defined here below 3 cycles per pupil) would be allocated into defocus, which can be compensated for during alignment. Therefore, we assumed that 8\,nm (half of 15.9) would correspond to defocus. The remaining surface error for each optic therefore becomes 9\,nm rms. 
Using numerical simulations, we estimated the scaled WFE contribution from each optic given the actual size of the beam compared to the two-inch optic. Under this simple assumption for mirrors O1, O2, O5, O6, and O9, the surface error becomes 6\,nm rms, and 5\,nm rms for O3, O4, O7, and O8. This conservative error budget gives a total WFE of 45\,nm rms prior to DM pre-compensation, assuming loose 31.7\,nm rms ($\lambda/20$ rms) alignment contribution to the WFE. 

Table \ref{table:optics_specs_for_manufacturers} also shows the as-built results. All optics surface errors of the manufactured parts are well within our specification range. 

\begin{figure}[!ht]
\centering
%\parbox{9cm}{
\includegraphics[scale=0.5]{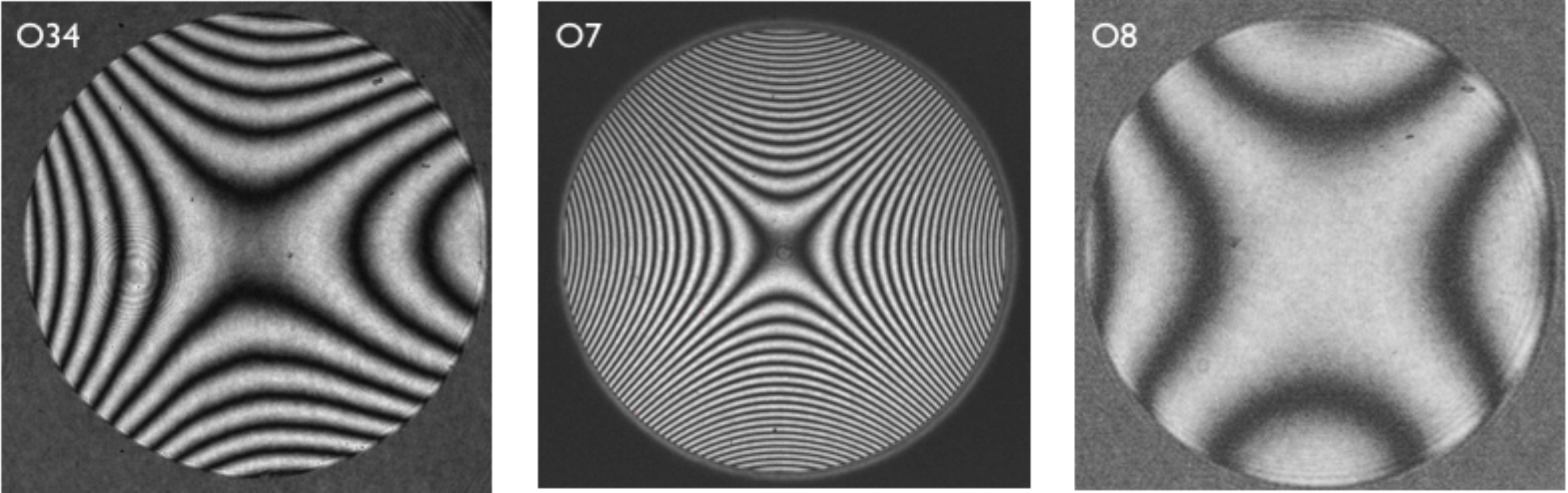}
%}
%\parbox{8cm}{
\caption{He-Ne interferogram of the three mirrors O34, O7 and O8 manufactured by the LAM for HICAT, underlining the super smooth quality of each mirror. As-built mid-spatial frequencies are below 2\,nm rms for all the mirrors, corresponding to 1.5 times smaller than our initial specification. In particular, the 6-inch parabola O34 was challenging to manufacture with f/1.4 ratio and the toric mirror O7 has $\sim 10$\,$\mu$m PV astigmatism.}
\label{fig:optics_results}
%}
\end{figure}

Our needs for high-contrast required particular attention to mid-spatial frequencies so we wanted to avoid small-tool machining for these optics. OAPs 1,2,5 and 6 were manufactured from a large mother parabola\footnote{http://www.nu-tek-optics.com} using large tools which typically lead to lower mid-spatial frequency errors. The most critical parts were manufactured by Laboratoire d'Astrophysique de Marseille\footnote{http://www.lam.fr}, following stress polishing methods for aspheric optics that have been experimented with success on the high-contrast imager VLT/SPHERE\cite{2012A&A...538A.139H}. Figure \ref{fig:optics_results} illustrates the results obtained for these optics, underlining the supersmooth quality of each mirror.

%_____________________________________________________________
\begin{table}[!ht]
\caption{Final optics specifications including a factor two surface error reserve margin for the manufacturers and as-built results. Wavefront errors are given for Low-, Mid-, and High-spatial (LoF, MiF, and HiF) frequency ranges.}
\centering
\begin{tabular}{l c c  | c c c |c c c | c}
\hline\hline
\multirow{3}{*}{\textbf{Optics}} & \textbf{Focal} & \textbf{WFE for} & \multicolumn{3}{c|}{\textbf{Final specifications}} & \multicolumn{3}{c|}{\textbf{As-built results}} & \textbf{Optic}\\
 & \textbf{length} & \textbf{alignment} & \multicolumn{3}{c|}{\textbf{surf. err. in nm rms}} & \multicolumn{3}{c|}{\textbf{surf. err. in nm rms}} & \textbf{shape}\\
 & \textbf{in mm} & \textbf{in nm rms} & \textbf{LoF} & \textbf{MiF} & \textbf{HiF} & \textbf{LoF} & \textbf{MiF} & \textbf{HiF} & \\
\hline
O2, O5, O6 & 478 & 31.7 &15.9 & 3.2 & 2.6 & $<$12.8 & $<$4.0 & $<$1.0 & Off-axis parabola\\
O3, O4 & 200 & 31.7 & 15.9 & 3.2 & 2.6 & $<$16 & 1.7 & $<$3 & one-piece parabola\\
O7 & 210 & 31.7  & 15.9 & 3.2 & 2.6 & 7 & 2 & $<$3 & Toric mirror\\
O8 & 713 & 31.7  & 15.9 & 3.2 & 2.6 & 7 & 1.5 & $<$3 & Toric mirror\\
O9 & 1597 & 31.7 & 15.9 & 3.2 & 2.6 & $<$11.4 & & & Spherical mirror\\
\hline
\end{tabular}\\
\label{table:optics_specs_for_manufacturers}
\end{table}
%_____________________________________________________________

%_____________________________________________________________
%_____________________________________________________________

%%%%%%%%%%%%%%%%%%%%%%%%%%%%%%%%%%%%%%%%%%%%%%%%%%%%%%%%%%%%%
\section{Opto-mechanical design}\label{sec:optomech}

\begin{figure}[!ht]
\centering
\includegraphics[width=\linewidth]{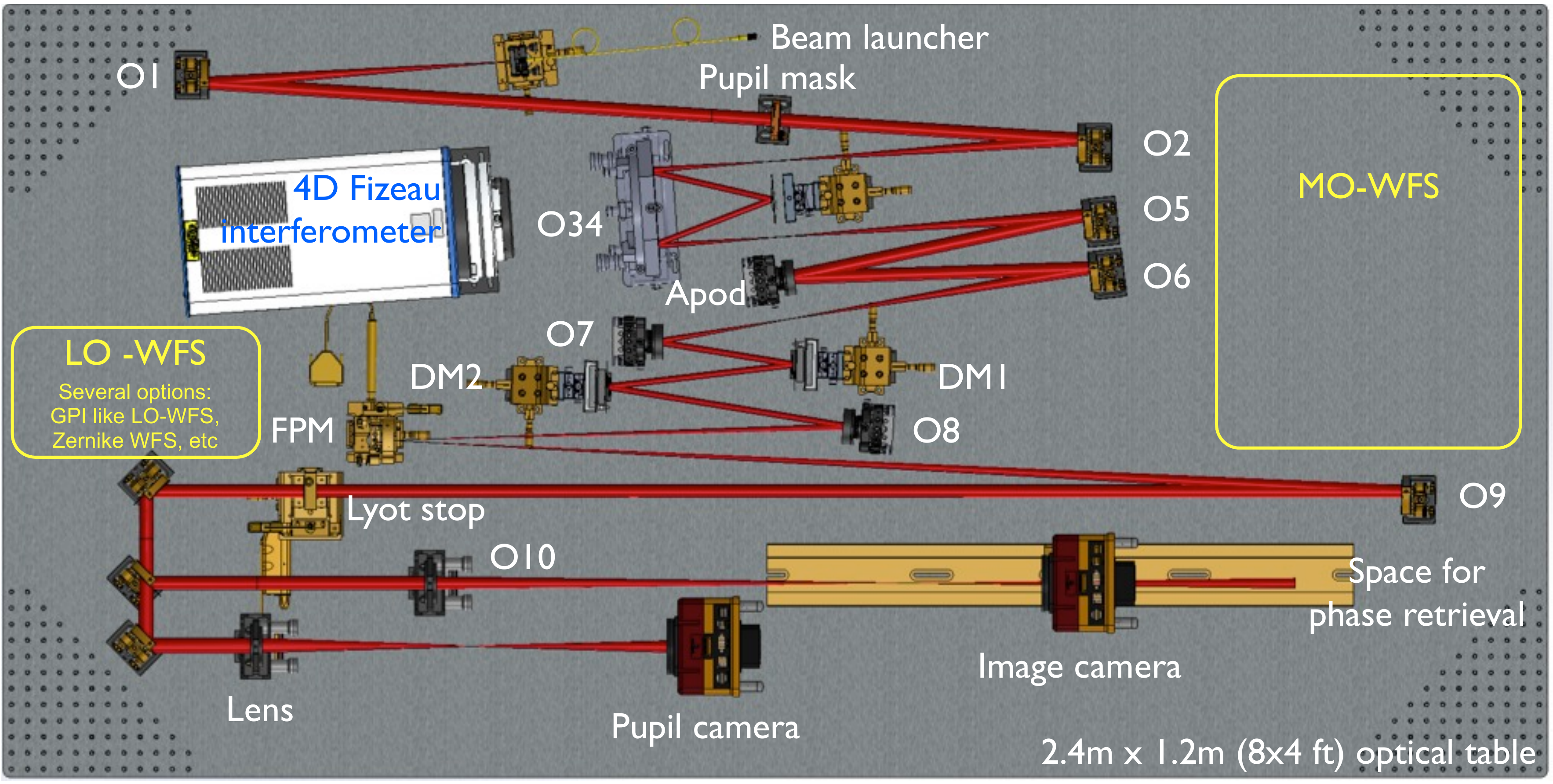}
\caption{Opto-mechanical model of the testbed, composed of nine reflective optics (from O1 to O9) and one refractive optic O10 to form the image on the camera. The optics O3 and O4 are part of a one-piece parabola. The reference flat is introduced in the optical path for use with the 4D AccuFiz interferometer. Beamsplitters (BS) are represented in the optical path for beam injection to or extraction from the main optical path. The image camera (CamF) is represented at an out-of-focus positions, showing flexibility for focal plane imaging and phase retrieval applications. The tested has also been designed to allow room for the implementation of a low-order wavefront sensor (LOWFS) behind the focal plane mask (FPM) such as Zernike wavefront sensors\cite{2011SPIE.8126E..11W,2012SPIE.8450E..0NN,2013A&A...555A..94N} and a mid- or high-order wavefront sensor (MO-WFS) using a possible pick-off close to the Lyot stop or using Lyot stop reflected light\cite{2010SPIE.7736E.179W,2009ApJ...693...75G}.}
\label{fig:testbed_optomech_design}
\end{figure}

Figure \ref{fig:testbed_optomech_design} shows the opto-mechanical model of the testbed, following the testbed design showed in HiCAT paper I\cite{2013SPIE.8864E..1KN}. The design includes space for a focal plane imager with motorization for phase diversity measurements, a pupil imager after the Lyot stop, and a 4D AccuFiz interferometer for alignments and direct wavefront sensing measurements. The current implementation accounts for two Boston kilo-DMs, which are expected in fall 2014, but with one engineering grade DM starting summer 2014. The apodizer has been chosen as a reflective optic to ease broadband operations. This part is initially replaced by a flat mirror but a design study has been carried out and detailed in N'Diaye et al. this conference\cite{2014ndiaye..aplc}.

The testbed is partially motorized. We set five actuators for the coronagraphic parts: three motors for positioning or not of the FPM into the beam and two actuators for the Lyot stop to allow clocking and introduction or not the diaphragm into the beam. The reflective mask and Lyot stop are planned to be operated remotely using a Labview interface. The other devices (beam launcher, segmented and continuous face sheet deformable mirrors) are manually adjustable with micrometer positioners with possible upgrade towards motors in the near future. 

Alignment tolerances for the different testbed optics have been established in our previous study\cite{2013SPIE.8864E..1KN} and based on their analysis, we select mounts for all the optics with tip tilt positioners to adjust them during the alignment process. Clocking positioning has been added to the apodizer for its alignment with the entrance pupil and segmented mirror, and to the toric mirrors O7 and O8 to account for their tight alignment tolerance for rotation around the optical axis. 

The opto-mechanical design was finalized in December 2013, with all parts delivered over the first quarter of 2014. Alignment work started early 2014.  

%_____________________________________________________________
%_____________________________________________________________

%%%%%%%%%%%%%%%%%%%%%%%%%%%%%%%%%%%%%%%%%%%%%%%%%%%%%%%%%%%%%
\section{Testbed alignment and first results}\label{sec:first_light}

\subsection{Alignments}

HiCAT presents more than 15 components on the table, which led us to develop an detailed alignment plan to position all the testbed optics accurately. Our goal is to minimize low-order wavefront errors due to misalignment errors to produce a source image with the best Strehl ratio at the coronagraph stage in the absence of DM-based wavefront control and wavefront shaping. 

As a first step, we sequentially align all the parts from O2 to O9 using our 4D AccuFiz interferometer as a beam launcher (instead of the source fiber and the optic O1) for beam injection into the design and direct wavefront sensor for aberration estimates. A given optic is finely tuned so as to reduce the low-order aberrations (power, both astigmatisms) observed with the interferometer. As a second step, we similarly align O1 by configuring the interferometer to inject light back from O9 to O2 and then, we position the fiber source at O1 focus to maximize the flux at the other side of the fiber which is connected to a power meter. We finally remove the 4D interferometer and launch the beam with the source fiber towards the Lyot stop to align this diaphragm, the fold mirrors, the lens O10 and the camera. We will eventually validate the end-to-end WFE using phase retrival (not implemented yet).

At the time of this communication, the reflective apodizer and the different deformable mirrors are not available yet. They are replaced by flat mirrors with excellent surface error quality ($\sim \lambda/20$ PV surface error over a 2-inch diameter). 
The pupil is a circular pupil mask with 18\,mm diameter instead of 20\,mm to account for initial potential distorsion or misalignments on the different pupil planes during initial alignment. Feedback from these analysis will help us to design second-generation pupil masks that provide low distorted beam at the different pupil planes.

\begin{figure}[!ht]
\centering
\includegraphics[width=0.75\linewidth]{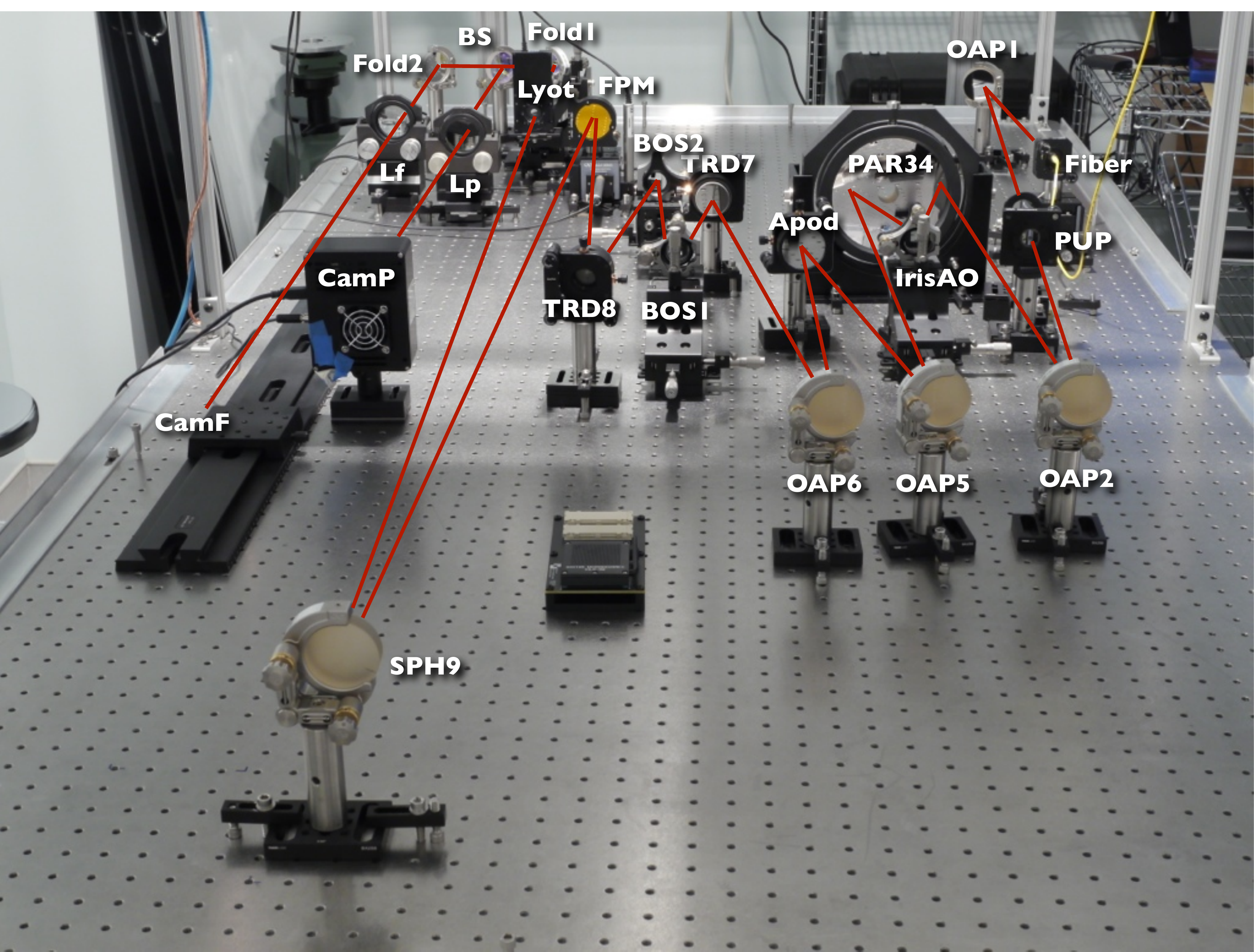}
\caption{Overall picture of the HiCAT testbed. The alignment was completed in June 2014.}
\label{fig:testbed_picture}
\end{figure}

\begin{figure}[!ht]
\centering
\includegraphics[width=0.75\linewidth]{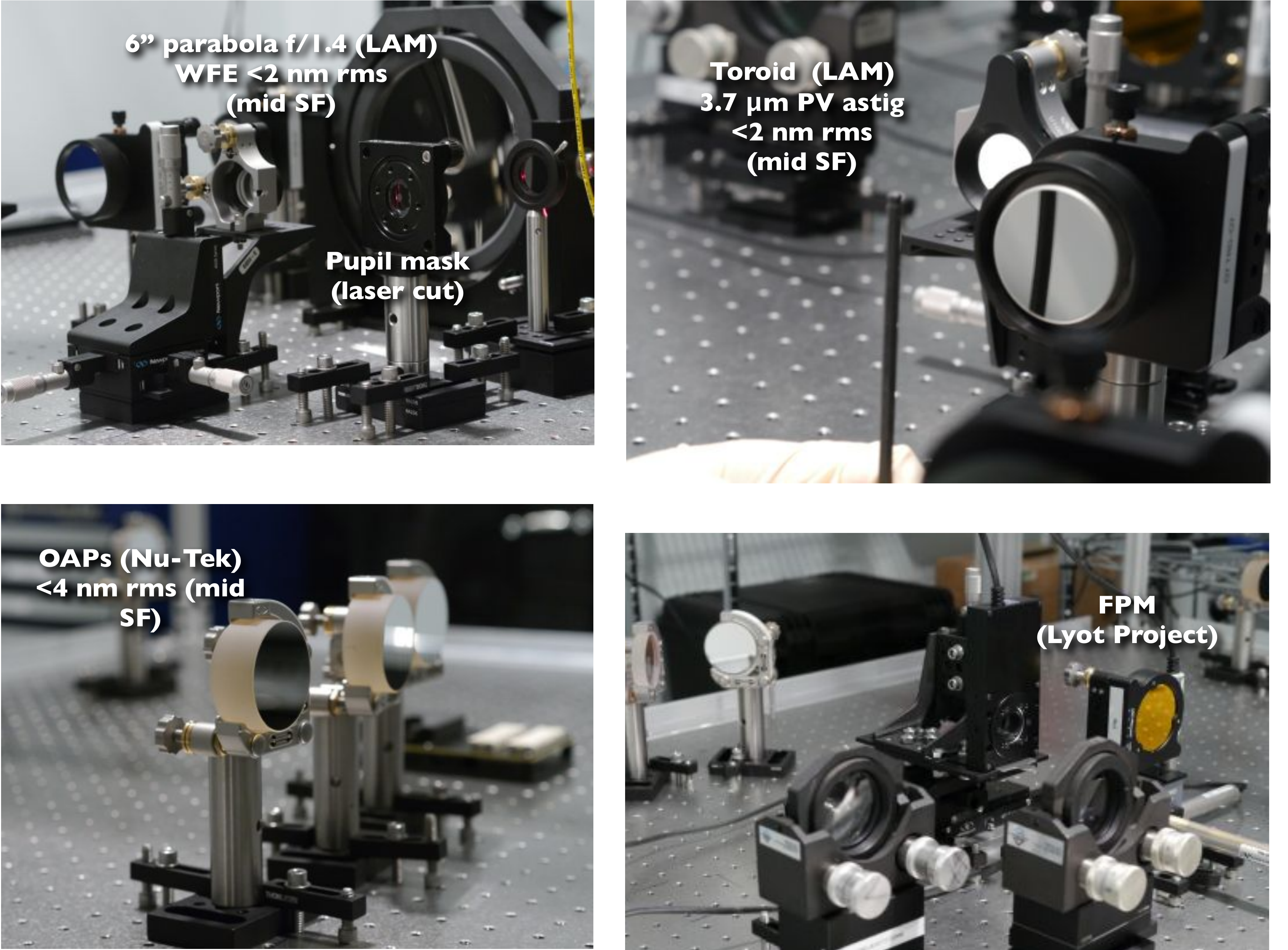}
\caption{Pictures of some HiCAT optics. \textbf{Top left}: the 6", f/1.4 ratio parabola manufactured by the LAM. A final WFE below 2nm rms in the mid-spatial frequencies was achieved for this extremely challenging optic to manufacture. \textbf{Top right}: one of the toric mirror manufactured by the LAM with a 3.7\,$\mu$m rms astigmatism. Here again, the WFE is below 2\,nm rms in the mid-spatial frequency range. \textbf{Bottom left}: the OAPs O1, O2, O5 and O6 manufactured by Nu-tek with a 4\,nm rms WFE in the mid-spatial frequencies. \textbf{Bottom right}: the gold-coated,  reflective FPM from the Lyot project.}
\label{fig:optics_picture}
\end{figure}

Figure \ref{fig:testbed_picture} shows a picture of the testbed in June 2014 after alignment of all the components. A wavefront error of 12\,$\pm$3nm rms has been measured from O1 to the FPM. This WFE is significantly better (by a factor 3 from our initial error budget) and due in part to the higher quality of the optics (see Figure \ref{fig:optics_picture}), very fine alignment achieved using 4D interferometer, and that most of the low-order aberrations consist of defocus compensated by the alignment.

%_____________________________________________________________
%_____________________________________________________________

%_____________________________________________________________
%_____________________________________________________________

\subsection{First results}

\subsubsection{Coronagraph mask focal plane}

\begin{figure}[!ht]
\centering
\includegraphics[width=0.75\linewidth]{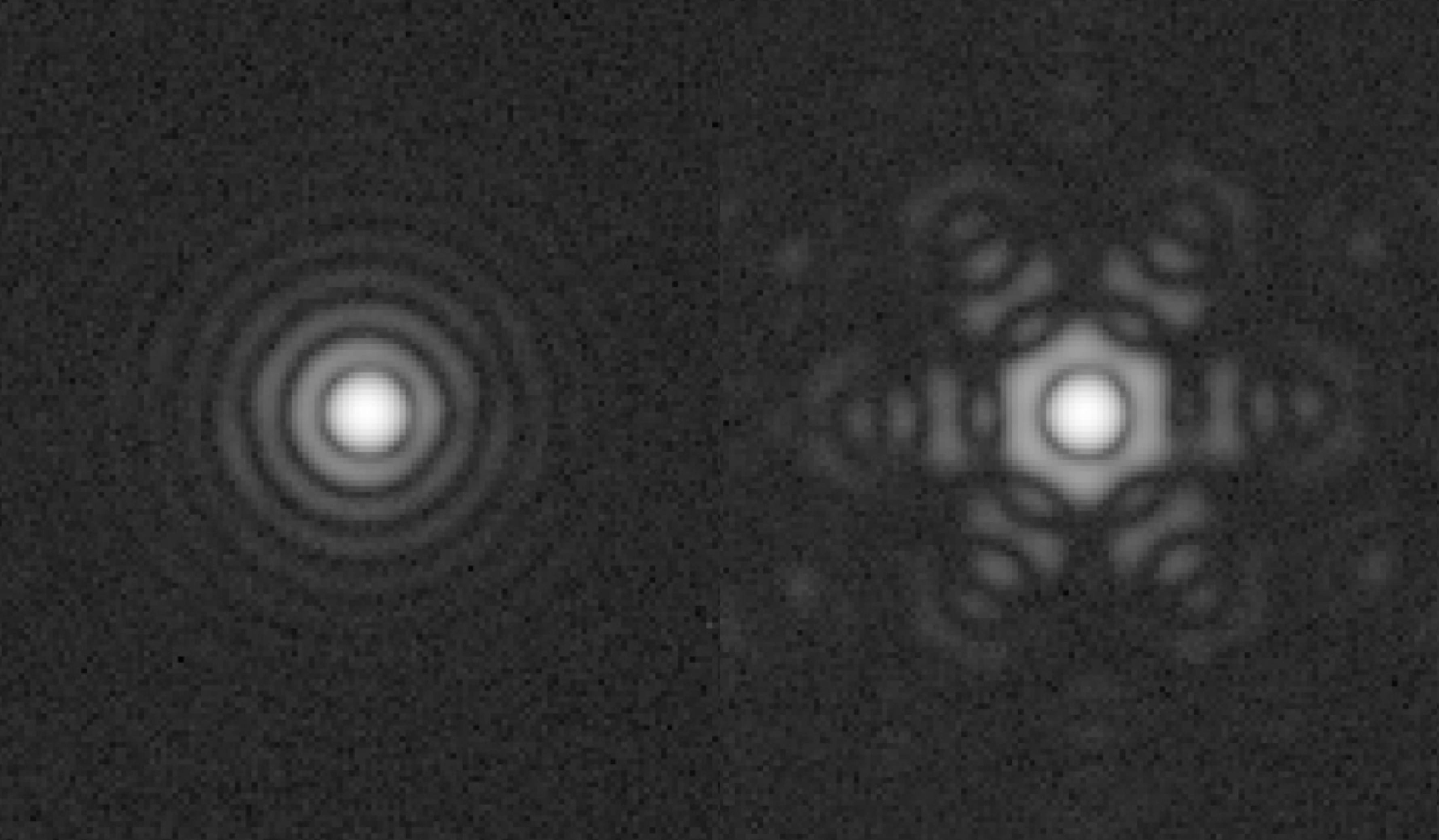}
\caption{PSFs in log scale captured at the level of the FPM for two different aperture shapes: a 18\,mm unobstructed circular aperture (left) and a JWST-like pupil (right) that was borrowed from the JWST Optical Simulation Testbed, see papers by Perrin et al. and Choquet et al. this conference. Seven regular diffraction rings can be observed in the left image, confirming the good quality of our alignment and our optics.}
\label{fig:psfs_FPM}
\end{figure}

After the alignment procedure, we produce images of our source at the level of the focal plane mask for a clear circular aperture and a JWST-like shaped pupil with central obstruction and spiders, see Figure \ref{fig:psfs_FPM}. Qualitatively, the PSF for the clear circular aperture presents annular rings with low distortion for the clear circular aperture, confirming the small amount of wavefront errors present in the system that was previously underlined with the interferometer wavefront sensing measurements ($\sim$12\,nm rms wavefront errors at FPM for a 18\,mm entrance pupil). 

%_____________________________________________________________
%_____________________________________________________________

\subsubsection{Lyot plane}

\begin{figure}[!ht]
\centering
\includegraphics[width=0.45\linewidth]{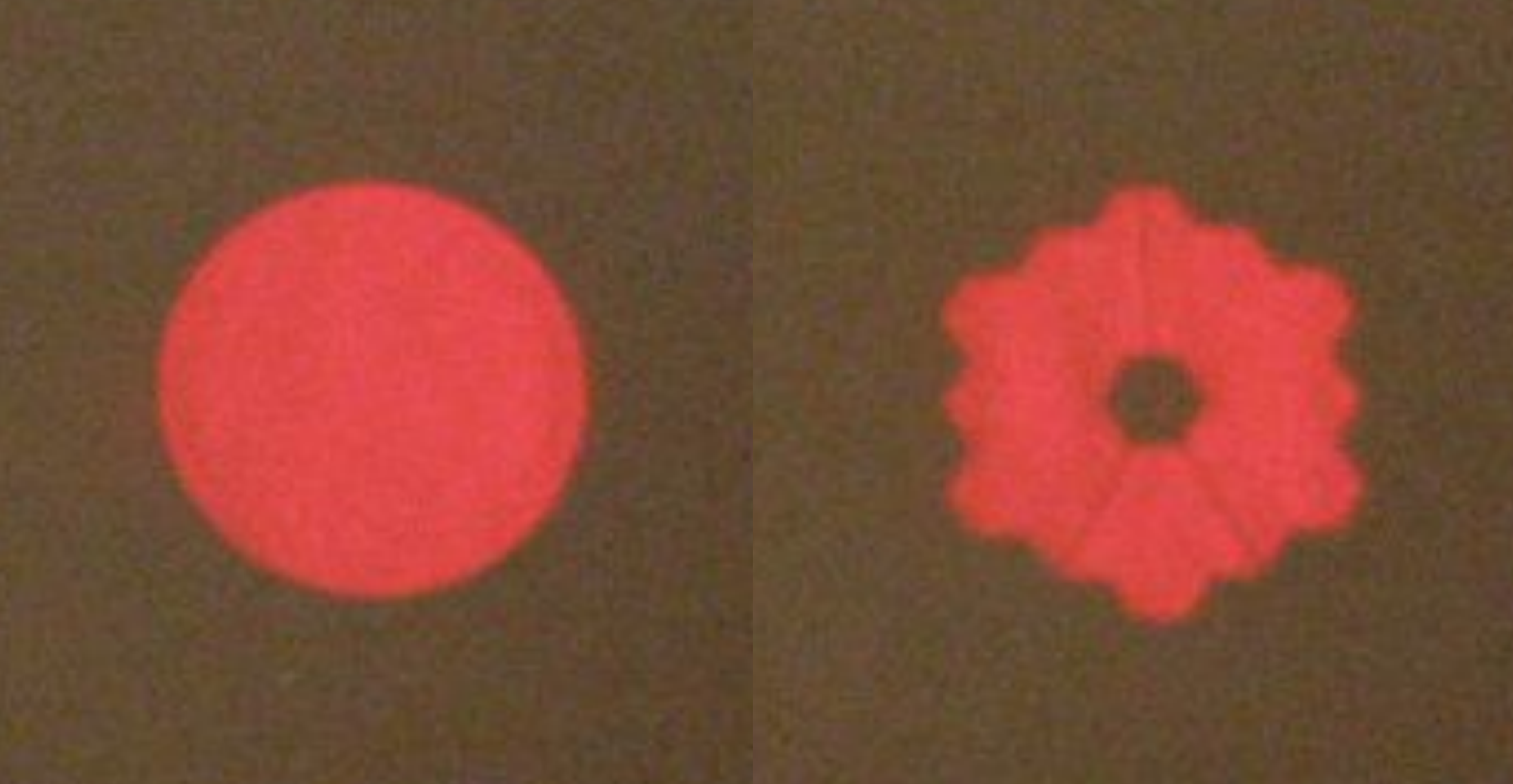}\\
\includegraphics[width=0.45\linewidth]{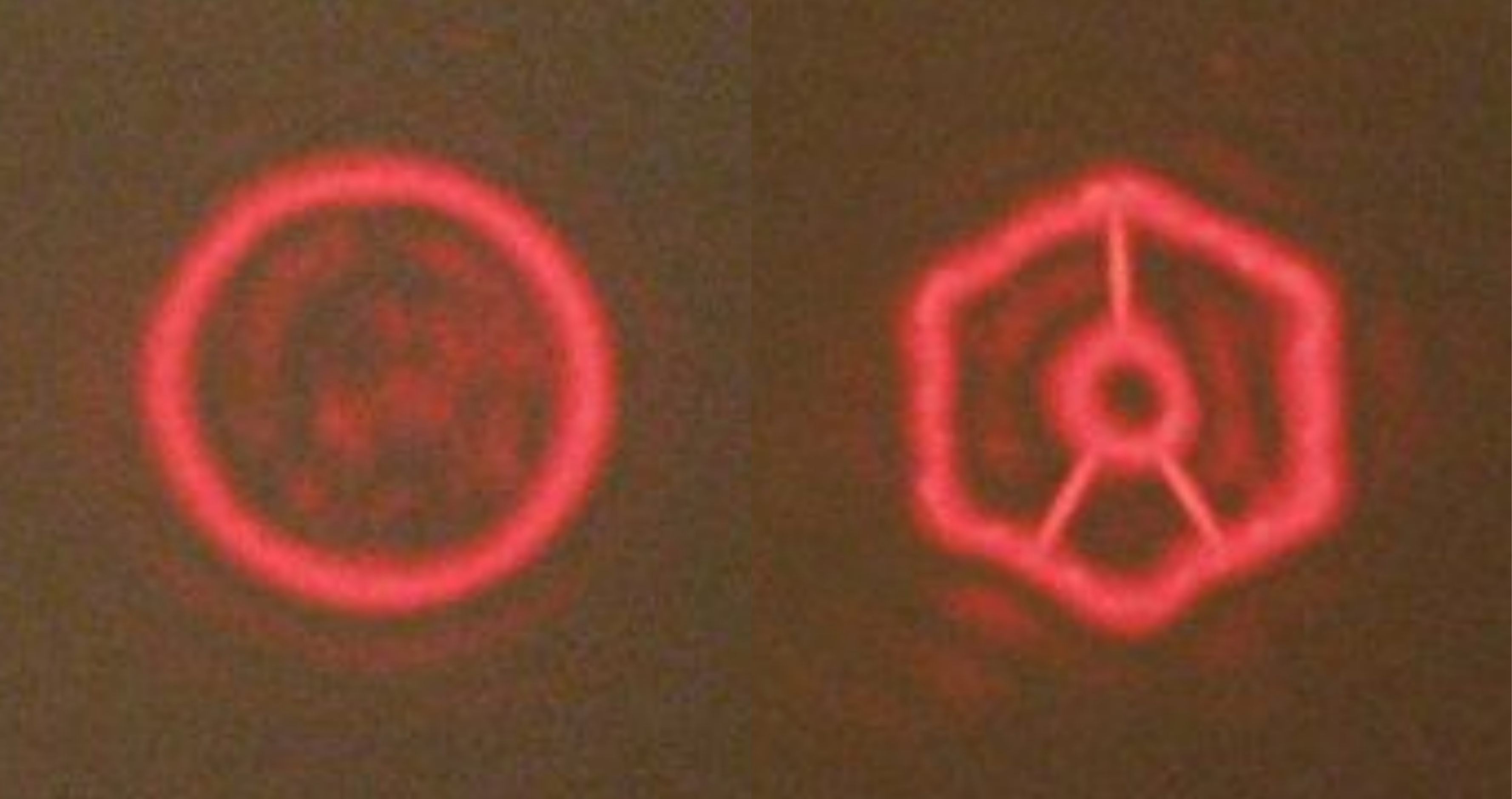}
\caption{Direct (top) and coronagraphic (bottom) images in the Lyot plane before stop application at $\lambda=$640\,nm in the presence of a clear circular aperture (left) and a JWST-like aperture (right). In the absence of pupil imager (CamP), we observe our first coronagraphic image in the relayed pupil using a screen.}
\label{fig:lyot_plane_image}
\end{figure}

Following these encouraging results, we generated coronagraphic images both in the Lyot plane and the camera image plane. We remind the reader that the apodizer is not available yet and replaced by a flat mirror. Also, the reflective FPM was not motorized yet and it was roughly centered onto the beam using manual positioners, allowing part of the beam to go through the 335\,$\mu$m diameter hole. Figure \ref{fig:lyot_plane_image} displays the first Lyot coronagraph images in the relayed pupil plane before Lyot stop application. The residual light is concentrated at the discontinuities of the aperture due to the removal of the PSF low-spatial frequencies by the FPM. Rings are also observable outside the geometric pupil, corresponding to the light diffracted by the FPM. We can also distinguish speckles within the geometric pupil, resulting from the residual wavefront errors that are not filtered by the coronagraph.

%_____________________________________________________________
%_____________________________________________________________

\subsubsection{Camera plane}

\begin{figure}[!ht]
\centering
\includegraphics[width=0.75\linewidth]{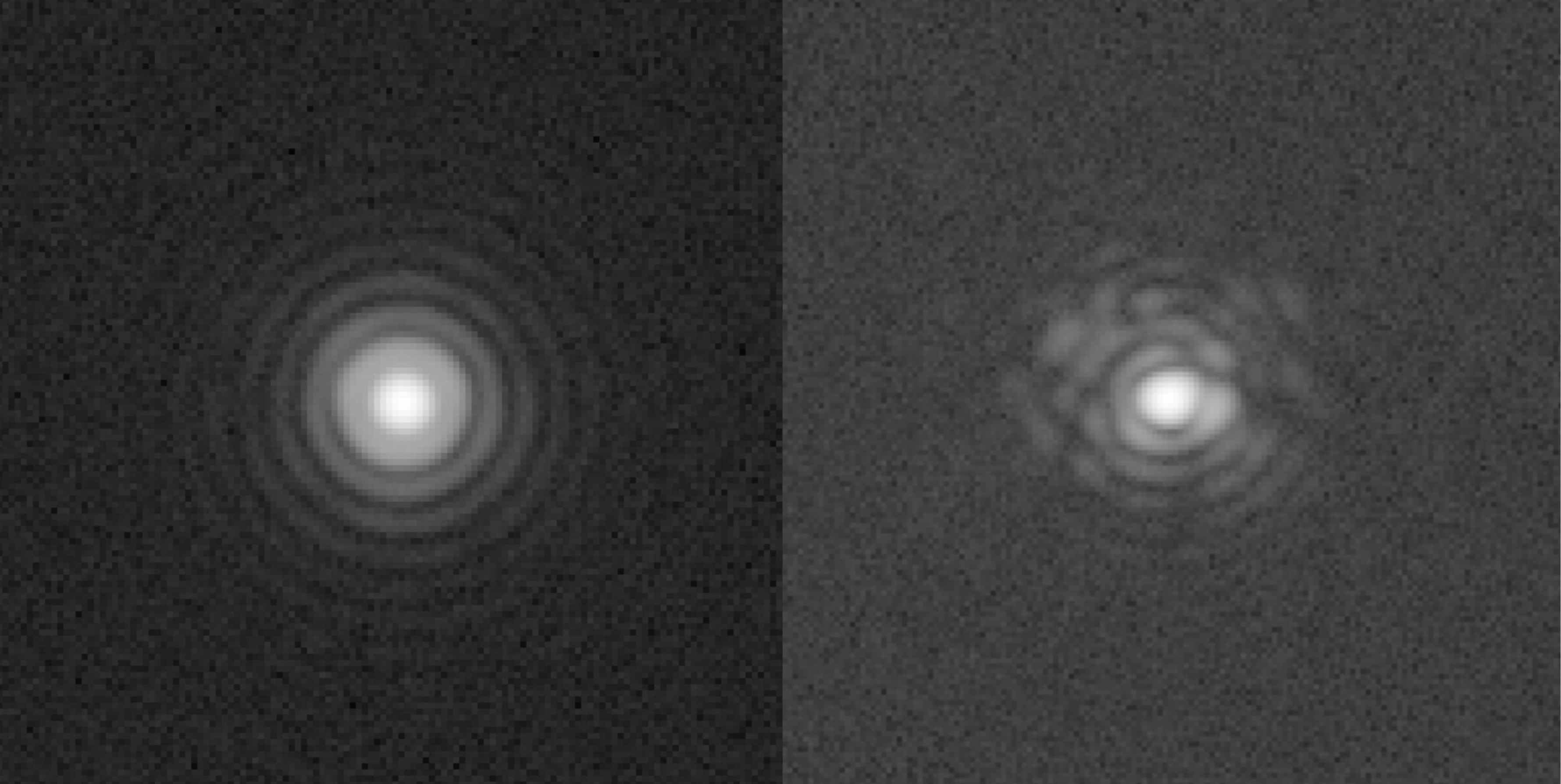}
\caption{Direct and coronagraphic images obtained on the HICAT testbed with a 18\,mm circular aperture and a 10\,mm Lyot stop. Both images are represented in log scale but not in the same scale. They have been taken in monochromatic light at $\lambda=640\,$nm.}
\label{fig:psfs_corono}
\end{figure}

We then focus our primary and qualitative studies on the 18\,mm circular aperture. The pupil at the Lyot plane has the same size as the entrance pupil mask. Therefore we introduce a 10\,mm circular Lyot stop in the relayed pupil plane to block the light diffracted by the FPM and we install the lens L10 to form our images in the camera image plane. Our first direct and coronagraphic images are displayed in Figure \ref{fig:psfs_corono}. While the direct image makes appear a nice Airy diffraction pattern, the coronagraphic image is filled with speckles resulting from the residual wavefront errors that have already been observed in the Lyot plane, as expected. These images are still preliminary with the FPM and Lyot that are possibly not optimally centered, and we did not make any contrast estimate before and after the coronagraph. However, these images are encouraging at qualitative level and recall the importance of controlling wavefront errors at the FPM very accurately to reduce the speckles in the coronagraphic field of view.

%_____________________________________________________________
%_____________________________________________________________

%%%%%%%%%%%%%%%%%%%%%%%%%%%%%%%%%%%%%%%%%%%%%%%%%%%%%%%%%%%%%
\section{Conclusions}

We have reported on the development of the HiCAT, the STScI high-contrast imaging testbed designed to provide a complete solution combining wavefront sensing and control, and starlight suppression strategies on telescopes with complex aperture geometries. We used an hybrid approach to realize the testbed optical design and define the surface error specifications of the optics that minimize the amplitude-induced errors, so-called Talbot effects, to develop methods for aperture control. 

Following our design studies, we provided our optic requirements to manufacturers and the as-built results are well within our specifications. We realized the opto-mechanical design and assembly of the testbed. After alignment of the different parts, we estimated a wavefront error of 12\,$\pm 3$nm rms at the level of the coronagraphic mask for a 18\,mm entrance pupil.

% resulting in one order of magnitude better than our budgeted wavefront error with our simulations for one-DM wavefront control and shaping. 

First direct and coronagraphic images have been obtained at different stages of the testbed (focal plane mask, Lyot plane, camera plane), qualitatively confirming the small amount of aberrations present in our system and validating our design studies. These preliminary results have been obtained in the absence of deformable mirrors (Iris AO segmented DM, kilo-DM Boston) and apodizer for the Lyot coronagraph. 

A first engineering grade Boston kilo-DM will be receive by the end of June and the two final science-grade DMs will be implemented in the fall 2014. Wavefront sensing methods, wavefront control and shaping strategies using the focal plane camera and one DM will be implemented with the Lyot coronagraph without apodizer during the second semester of 2014 to produce a PSF dark hole at a moderate contrast during the Fall/Winter 2014.

Coronagraph optimizations including apodizations are currently being developed in the context of HiCAT to have access to $10^8$ contrast levels, see our first results in N'Diaye et al. this conference\cite{2014ndiaye..aplc}.

%%%%%%%%%%%%%%%%%%%%%%%%%%%%%%%%%%%%%%%%%%%%%%%%%%%%%%%%%%%%%
\acknowledgments     %>>>> equivalent to \section*{ACKNOWLEDGMENTS}       
This work is supported by the National Aeronautics and Space Administration under Grant NNX12AG05G issued through the Astrophysics Research and Analysis (APRA) program (PI: R. Soummer). The authors warmly acknowledge Alexis Carlotti, Tyler Groff, George Hartig, N. Jeremy Kasdin, Matt Kenworthy, Charles-Philippe Lajoie, Bruce Macintosh, Dimitri Mawet, Colin Norman, Matt Sheckells, and Stuart Shaklan for fruitful discussions during the testbed design. The authors are also grateful to Nicole Cade-Ferreira, Andrew Frazier, Matthew Jorgensen, and Tucker Kearney from the Department of the Mechanical Engineering of Johns Hopkins University, for the design, development and integration of the HiCAT testbed enclosure.

L.P., M.P.D., and R.S. conceived of this project and led the overall effort. M.N. and E.C. developed the whole design and simulation studies for the testbed. O.L. realized the opto-mechanical design. S.E. implemented the optical and mechanical design, and with L.L., they performed the fine alignment of the testbed and capture images. E.E. helped with the optical design and the tolerancing studies. A.D. performed the drawings of the optics for the manufacturers. J.K.W. provided valuable advice and guidance for the alignment. C.L. helped with assembly and machining. E.H., M.M. and M.F. developed and manufactured the 6-inch parabola and the toric mirrors and they also contributed to the optical design. As lab manager R.A. coordinated support and lab infrastructure for the new testbed.

%%%%%%%%%%%%%%%%%%%%%%%%%%%%%%%%%%%%%%%%%%%%%%%%%%%%%%%%%%%%%
%%%%% References %%%%%
\bibliography{2014_mndiaye_spie}   %>>>> bibliography data in report.bib
\bibliographystyle{spiebib}   %>>>> makes bibtex use spiebib.bst

\end{document}